\documentclass[copyright,creativecommons]{eptcs}
\providecommand{\event}{TFPIE 2023} 

\usepackage{iftex}
\usepackage{underscore}         
\usepackage[T1]{fontenc}        
\usepackage[utf8]{inputenc}
\usepackage{stmaryrd}
\usepackage{amsmath}
\usepackage{amssymb}
\usepackage{xspace}
\usepackage{prettyref}
\usepackage{xcolor}
\usepackage[frozencache,cachedir=minted-cache]{minted}
\usepackage[all]{xy}
\usepackage{amsfonts}
\usepackage{graphicx}
\usepackage{hyperref}
\usepackage{array}

\newcommand{\disco}{\textsc{Disco}\xspace}

\newcommand{\pkg}[1]{\href{https://hackage.haskell.org/package/#1}{\texttt{#1}}}

\newcommand{\ie}{\emph{i.e.}\ }
\newcommand{\eg}{\emph{e.g.}\ }
\newcommand{\etal}{\emph{et al.}\xspace}

\newrefformat{fig}{Fig.~\ref{#1}}
\newrefformat{sec}{Section~\ref{#1}}
\newrefformat{eq}{Equation~\eqref{#1}}
\newrefformat{prob}{Problem~\ref{#1}}
\newrefformat{tab}{Table~\ref{#1}}
\newrefformat{thm}{Theorem~\ref{#1}}
\newrefformat{lem}{Lemma~\ref{#1}}
\newrefformat{prop}{Proposition~\ref{#1}}
\newrefformat{defn}{Definition~\ref{#1}}
\newrefformat{cor}{Corollary~\ref{#1}}
\newrefformat{lst}{Listing~\ref{#1}}
\newcommand{\pref}[1]{\prettyref{#1}}

\newif\ifcomments\commentsfalse

\ifcomments
\newcommand{\authornote}[3]{\textcolor{#1}{[#3 ---#2]}}
\newcommand{\todo}[1]{\textcolor{red}{[TODO: #1]}}
\else
\newcommand{\authornote}[3]{}
\newcommand{\todo}[1]{}
\fi

\newcommand{\N}{\mathbb{N}}
\newcommand{\Z}{\mathbb{Z}}
\newcommand{\F}{\mathbb{F}}
\newcommand{\Q}{\mathbb{Q}}

\DeclareUnicodeCharacter{27C5}{$\Lbag$}
\DeclareUnicodeCharacter{27C6}{$\Rbag$}
\DeclareUnicodeCharacter{2192}{$\to$}
\DeclareUnicodeCharacter{D7}{$\times$}
\DeclareUnicodeCharacter{2200}{$\forall$}
\DeclareUnicodeCharacter{2115}{$\N$}
\DeclareUnicodeCharacter{2124}{$\Z$}
\DeclareUnicodeCharacter{211A}{$\Q$}
\DeclareUnicodeCharacter{1D53D}{$\F$}
\DeclareUnicodeCharacter{03BB}{$\lambda$}

\title{\disco: A Functional Programming Language for Discrete Mathematics}
\author{Brent A. Yorgey
\institute{Hendrix College\\ Conway, Arkansas, USA}
\email{yorgey@hendrix.edu}
}
\def\titlerunning{\disco: A Functional Programming Language for Discrete Mathematics}
\def\authorrunning{B. A. Yorgey}

\begin{document}
\maketitle


\begin{abstract}
  \disco is a pure, strict, statically typed functional programming
  language designed to be used in the setting of a discrete
  mathematics course. The goals of the language are to introduce
  students to functional programming concepts early, and to enhance
  their learning of mathematics by providing a computational platform
  for them to play with.  It features mathematically-inspired notation,
  property-based testing, equirecursive algebraic types, subtyping,
  built-in list, bag, and finite set types, a REPL, and student-focused
  documentation.  \disco is implemented in Haskell, with source code
  available on
  GitHub,\footnote{\url{https://github.com/disco-lang/disco}} and
  interactive web-based REPL available through
  replit.\footnote{\url{https://replit.com/@BrentYorgey/Disco\#README.md}}
\end{abstract}

\section{Introduction}
\label{sec:introduction}

Many computer science curricula at the university level include a
\emph{discrete mathematics} course as a core requirement
\cite{ACM:2013}.  Often taken in the first or second year, a discrete
mathematics course introduces mathematical structures and techniques
of foundational importance in computer science, such as induction and
recursion, set theory, logic, modular arithmetic, functions,
relations, and graphs.  In addition, it sometimes serves as an
introduction to writing formal proofs.  Although there is wide
agreement that discrete mathematics is foundational, students often
struggle to see its relevance to computer science.

\emph{Functional programming} is a style of programming, embodied in
languages such as Haskell, Standard ML, OCaml, Scala, F\#, and Racket, which
emphasizes functions (\ie input-output processes) rather than
sequences of instructions. It enables working at high levels of
abstraction as well as rapid prototyping and refactoring, and provides
a concise and powerful vocabulary to talk about many other topics in
computer science.  It is becoming critical to expose undergraduate
students to functional programming early, but many computer science
programs struggle to make space for it.  The Association for Computing
Machinery's 2013 curricular guidelines \cite{ACM:2013} do not even
include functional programming as a core topic.

One creative idea is to combine functional programming and discrete
mathematics into a single course.  This is not a new idea
\cite{Wainwright:1992, Henderson:2002, Scharff:2002, Doets:2004,
  ODonnell:2006, VanDrunen:2011, vandrunen2013discrete,
  10.1145/3078325, Xing:2008}, and even shows up
in the 2007 model curriculum of the Liberal Arts Computer Science
Consortium \cite{LiberalArtsComputerScienceConsortium:2007}. The
benefits of such an approach are numerous:
\begin{itemize}
\item It allows functional programming to be introduced at an early
  point in undergraduates' careers, since discrete mathematics is
  typically taken in the first or second year.  This allows ideas from
  functional programming to inform students' thinking about the rest
  of the curriculum.  By contrast, when functional programming is left
  until later in the course of study, it is in danger of being seen as
  esoteric or as a mere curiosity.
\item The two subjects complement each other well: discrete math
  topics make good functional programming exercises, and ideas from
  functional programming help illuminate topics in discrete mathematics.
\item In a discrete mathematics course with both mathematics and
  computer science majors, mathematics majors can have a ``home turf
  advantage'' since the course deals with topics that may be already
  familiar to them (such as writing proofs), whereas computer science
  majors may struggle to connect the course content to computer
  science skills and concepts they already know.  Including functional
  programming levels the playing field, giving both groups of students
  a way to connect the course content to their previous experience.
  Computer science majors will be more comfortable learning
  mathematical concepts that they can play with computationally;
  mathematics majors can leverage their experience with mathematics to
  learn a bit about programming.
\item It is just plain fun: using programming enables interactive
  exploration of mathematical concepts, which leads to higher
  engagement and increased retention.
\end{itemize}

However, despite its benefits, this model is not widespread in
practice.  This may be due partly to lack of awareness, but there are
also some real roadblocks to adoption that make it impractical or
impossible for many departments.

\begin{itemize}
\item Existing functional languages---such as Haskell, Racket, OCaml,
  or SML---are general-purpose languages which (with the notable
  exception of Racket) were not designed specifically with teaching in
  mind.  The majority of their features are not needed in the setting
  of discrete mathematics, and teachers must waste a lot of time and
  energy explaining incidental detail or trying to hide it from
  students.
\item Again with the notable exception of Racket, tooling for existing
  functional languages is designed for professional programmers, not
  for students.  The systems can be difficult to set up, generate
  confusing error messages, and are generally designed to facilitate
  efficient production of code rather than interactive exploration and
  learning.
\item As with any subject, effective teaching of a functional language
  requires expertise in the language and its use, or at least thorough
  familiarity, on the part of the instructor. General-purpose
  functional languages are large, complex systems, requiring deep
  study and years of experience to master.  Even if only a small part
  of the language is presented to students, a high level of expertise
  is still required to be able to select and present a relevant subset
  of the language and to help students navigate around the features
  they do not need.  For many instructors, spending years learning a
  general-purpose functional language just to teach discrete
  mathematics is a non-starter.  This is especially a problem at
  institutions where the discrete mathematics course is taught by
  mathematics faculty rather than computer science faculty.
\item Students often experience friction caused by differences between
  standard mathematics notation and the notation used by existing
  functional programming languages.  As one simple example, in
  mathematics one can write $2x$ to denote multiplication of $x$ by
  $2$; but many programming languages require writing a multiplication
  operator, for example, \texttt{2*x}. Any one such difference is
  small, but the accumulation of many such differences can be a real
  impediment to students as they attempt to move back and forth
  between the worlds of abstract mathematics and concrete computer
  programs.  For example, consider the following function defined
  using typical mathematical notation:
  \begin{align*}
    f &: \N \to \Q \\
    f(2n)   &= 0 \\
    f(2n+1) &= \begin{cases} n/2 & \text{if } n > 5, \\
      3n + 7 & \text{otherwise}
      \end{cases}
  \end{align*}
  Now consider this translation of the function into idiomatic
  Haskell:
  \begin{minted}{haskell}
f :: Int -> Rational
f x
  | even x    = 0
  | n > 5     = fromIntegral n / 2
  | otherwise = 3*n + 7
  where
    n = x `div` 2
  \end{minted}
  Although the translation may seem trivial to experienced functional
  programmers, from the point of view of a student these are extremely
  different.
\end{itemize}

\disco is a new functional programming language, specifically designed
for use in a discrete mathematics course, which attempts to solve many
of these issues:

\begin{itemize}
\item Although \disco is Turing-complete, it is a teaching language,
  not a general-purpose language.  It includes only features which are
  of direct relevance to teaching core functional programming and
  discrete mathematics topics; for example, it does not include a
  floating-point number type.  \pref{sec:examples} has many examples
  of the language's features and some discussion of features which are
  explicitly excluded.
\item As much as possible, the language's features and syntax mirror
  common mathematical practice rather than other functional
  languages. For example, a translation into \disco of the example
  function introduced previously is shown below.
    \begin{minted}{text}
f : N -> Q
f(2n)   = 0
f(2n+1) = {? n/2      if n > 5,
             3n + 7   otherwise
          ?}
  \end{minted}
  \pref{sec:examples}
  has many more examples, and \pref{sec:syntax} discusses some notable
  exceptions.
\item As a result---although there is as yet no data to back this
  up---the language should be easy for instructors to learn, even
  mathematicians without much prior programming experience.
\end{itemize}

\disco is an open-source project, implemented in Haskell, with source
code licensed under a BSD 3-clause license and available on
GitHub.\footnote{\url{https://github.com/disco-lang/disco}} Although
it is possible to install \disco locally, either from
Hackage\footnote{\url{https://hackage.haskell.org/package/disco}} or
directly from source, one can also interact with \disco in the cloud
via a web browser, through the magic of
replit.\footnote{\url{https://replit.com/@BrentYorgey/Disco\#README.md}}
This is the primary way that students will be instructed to use Disco,
so that students do not need to install a Haskell toolchain or worry
about exhausting the computational resources of their device.  Via
replit, it is entirely feasible to play with \disco on any device with
a web browser, including Chromebooks, tablets, or phones.
Documentation for \disco is hosted on
\texttt{readthedocs.org}.\footnote{\url{https://disco-lang.readthedocs.io}}


\section{\disco by Example}
\label{sec:examples}

We will begin by exploring some of the major features and uses of the
language via a series of examples.

\subsection{Greatest common divisor}
\label{sec:gcd}

Our first example is an implementation of the classic Euclidean
algorithm for computing the greatest common divisor of two natural
numbers, shown in \pref{lst:gcd}.

\begin{listing}[!htp]
\inputminted{text}{examples/gcd.disco}
\caption{Definition of \texttt{gcd} in \disco}
\label{lst:gcd}
\end{listing}

Lines beginning with \texttt{|||} denote special documentation
comments attached to the subsequent definition, similar to docstrings
in Python (regular comments start with \texttt{-{}-}).  This
documentation can be later accessed with the \texttt{:doc} command at
the REPL prompt:

\begin{verbatim}
Disco> :doc gcd
gcd : ℕ × ℕ → ℕ

The greatest common divisor of two natural numbers.

\end{verbatim}

Lines beginning with \texttt{!!!} denote \emph{tests} attached to the
subsequent definition, which can be either simple Boolean unit tests
(such as \verb|gcd(7,6) == 1|), or quantified properties (such as the
last two tests, which together express the universal property defining
\verb|gcd|).  Such properties will be tested exhaustively when
feasible, or, when exhaustive testing is impossible (as in this case),
tested with a finite number of randomly chosen inputs. Under the hood,
this uses the \pkg{QuickCheck} \cite{claessen2000quickcheck} and
\pkg{simple-enumeration} packages to generate inputs.  For example:

\begin{verbatim}
Disco> :test forall a:N, b:N. let g = gcd(a,b) in g divides a /\ g divides b
  - Possibly true: ∀a, b. let g = gcd(a, b) in g divides a /\ g divides b
    Checked 100 possibilities without finding a counterexample.

Disco> :test forall a:N, b:N. let g = gcd(a,b) in g divides a /\ (2g) divides b
  - Certainly false: ∀a, b. let g = gcd(a, b) in g divides a /\ 2 * g divides b
    Counterexample:
      a = 0
      b = 1
\end{verbatim}

In the first case, \disco reports that 100 sample inputs were checked
without finding a counterexample, leading to the conclusion that the
property is \emph{possibly} true.  In the second case, when we modify
the test by demanding that \verb|b| must be divisible by twice
\verb|gcd(a,b)|, \disco is quickly able to find a counterexample,
proving that the property is \emph{certainly} false.

Every top-level definition in \disco must have a type signature;
\verb|gcd : N * N -> N| indicates that \verb|gcd| is a function which
takes a pair of natural numbers as input and produces a natural number
result.  The recursive definition of \verb|gcd| is then
straightforward, featuring multiple clauses and pattern-matching on
the input.

\subsection{Primality testing}
\label{sec:primetest}

The example shown in \pref{lst:prime}, testing natural numbers for
primality via trial division, is taken from Doets and van
Eijck~\cite[pp. 4--11]{Doets:2004}, and has been transcribed from
Haskell into \disco. (\disco also has a much more efficient built-in
primality testing function that calls out to the highly optimized
\pkg{arithmoi} package.)

\begin{listing}[!htp]
\inputminted{text}{examples/prime.disco}
\caption{Primality testing in \disco}
\label{lst:prime}
\end{listing}

There are a few interesting things to point out about this example.
The most obvious is the use of a \emph{case expression} in the
definition of \verb|ldf| delimited by \verb|{? ... ?}|.  It is
supposed to be reminiscent of typical mathematical notation like
\[ \mathit{ldf}\;k\;n = \begin{cases} k & \text{if } k \mid n, \\ n &
    \text{if } k^2 > n, \\ \mathit{ldf}\;(k+1)\;n &
    \text{otherwise.} \end{cases} \] However, we can't use a bare
curly brace as \disco syntax since it would conflict with the notation
for literal sets (and we can't use a giant, multi-line curly brace in
any case!\footnote{One might imagine using vertically aligned curly
  brace characters to simulate a giant curly brace, but that would require
  tricky indentation-sensitive parsing.}).  The intention is that writing %
\verb|{? ... ?}| lends itself to the mnemonic of ``asking questions''
to see which branch of the case expression to choose.  In general,
each branch can have multiple chained conditions, each of which can
either be a Boolean guard, as in this example, or a pattern match
introduced with the \verb|is| keyword.  In fact, all multi-clause
function definitions with pattern matching really desugar into a
single case expression. For example, the definition of \verb|gcd| in
\pref{lst:gcd} desugars to
\begin{minted}{text}
gcd : N * N -> N
gcd = \p. {? a if p is (a,0), gcd(b, a mod b) if p is (a,b) ?}
\end{minted}

Notice that the definition of \verb|isPrime| uses the \verb|and|
keyword instead of \verb|/\|.  These are synonymous---in fact,
\verb|&&| and $\land$ (U+2227 LOGICAL AND) are also accepted.  In
general, \disco's philosophy is to allow multiple syntaxes for things
with common synonyms rather than imposing one particular choice.
Typically, a Unicode representation of the ``real'' notation is
supported (and used when pretty-printing), along with an ASCII
equivalent, as well as (when applicable) syntax common in other
functional programming languages.  Another good example is the natural
number type, which can be written $\N$, \verb|N|, \verb|Nat|, or
\verb|Natural|.  There are several reasons for this design choice:
\begin{itemize}
\item It makes code easier to \emph{write} since students have to
  spend less time trying to remember the one and only correct syntax
  choice, or worrying about whether a particular syntax they remember
  comes from math class, Python, or \disco.
\item Although having many different syntax choices can make code
  harder to \emph{read}, helping students learn how to interpret
  formal notation and how to translate between mathematics and
  programming notation are typical explicit learning goals of the
  course, so this could be considered a feature.
\end{itemize}

Notice that \verb|ldf| is defined via currying, and is partially
applied in the definition of \verb|ld|. Just as in Haskell, every
function in \disco takes exactly one argument, but some
functions can return other functions (curried style) and some functions
can take a product type as input (uncurried style).  Via tutorials,
documentation, and the types of standard library functions, \disco
encourages the use of an \emph{uncurried} style, since students are
already used to notation like \verb|f(x,y)| for multi-argument
functions in mathematics.

Finally, this example introduces the primitive \verb|Bool| type in
addition to the natural number type \verb|N| seen previously.  \disco
also has a primitive \verb|Char| type for Unicode codepoints, and
several other numeric types to be discussed later.

\subsection{Z-order}
\label{sec:zorder}

The ``Morton Z-order'' is one of my favorite bijections showing that
$\N \times \N$ has the same cardinality as $\N$; it takes a pair of
natural numbers, expresses them in binary, and interleaves their
binary representations to form a single natural number.  \disco code
to compute this bijection (and check that it really is a bijection) is
shown in~\pref{lst:zorder}.

\begin{listing}[!htp]
\inputminted{text}{examples/zorder.disco}
\caption{Morton Z-Order}
\label{lst:zorder}
\end{listing}

This example again uses case expressions; it may seem odd to use case
expressions with only one branch, but this is done in order to be able
to pattern-match on the result of the recursive call to
\mintinline{text}{zOrder'}.  The most interesting thing about this
example is its use of \emph{arithmetic patterns}, such as %
\mintinline{text}{zOrder'(2n) = ...} and
\mintinline{text}{zOrder'(2n+1) = ...}.  This is common mathematical
notation, but perhaps less common in programming languages.  Any
expression with exactly one variable and only basic arithmetic
operators can be used as a pattern; the pattern matches if there
exists a value of the appropriate type for the variable which makes
the expression equal to the input.  For example, the pattern
\texttt{2n} will match only even natural numbers, and \texttt{n} will
then be bound to half of the input.

\subsection{Finite sets}

\disco has built-in \emph{finite sets}; in particular, values of type
\texttt{Set(A)} are finite sets with elements of type
\texttt{A}. \disco supports the usual set operations (union,
intersection, difference, cardinality, power set), and sets can be created by
writing a finite set literal, like \verb|{1,3,5,7}|, using ellipsis
notation, like \verb|{1, 3 .. 7}|, or using a set comprehension, as in
\verb-{2x+1 | x in {0 .. 3}}-. \pref{lst:sets} shows a portion
of an exercise (with answers filled in) to help students practice
their understanding of set comprehensions.

\begin{listing}[!htp]
\inputminted{text}{examples/sets.disco}
\caption{Set comprehension exercise}
\label{lst:sets}
\end{listing}

Set comprehensions in \disco work similarly to list comprehensions in
Haskell (\disco has list and bag comprehensions as well).  In
these examples we can see both \emph{filtering} the generated values via
Boolean guards, as well as \emph{transforming} the outputs via an
expression to the left of the vertical bar.

While on the subject of sets, it is worth mentioning that the
distinction between \emph{types} and \emph{sets} is something of a
pedagogical minefield: the distinction is nonexistent in typical
presentations of mathematics, but crucial in a computational system
with static type checking.  This issue is discussed in more detail in
\pref{sec:typesvsets}.

One other thing this example highlights is that there is extensive,
student-centered documentation available at
\url{https://disco-lang.readthedocs.io/}.  Students are pointed to
this documentation not just from links in homework assignments such as
this, but also by the \disco REPL itself. Encountering an error, or
asking for documentation about a function, type, or operator, are
all likely to result in documentation links for further reading, as
illustrated in \pref{lst:doc}.

\begin{listing}[!htp]
\begin{minted}{text}
Disco> :doc +
~+~ : ℕ × ℕ → ℕ
precedence level 7, left associative

The sum of two numbers, types, or graphs.

https://disco-lang.readthedocs.io/en/latest/reference/addition.html

Disco> x + 3
Error: there is nothing named x.
https://disco-lang.readthedocs.io/en/latest/reference/unbound.html
\end{minted}
\caption{\disco generates links to online documentation}
\label{lst:doc}
\end{listing}

\subsection{Trees and Catalan numbers}

\pref{lst:catalan} is a fun example generating and counting binary
trees. It defines a recursive type \texttt{BT} of binary tree shapes,
along with a function to generate a list of all possible tree shapes
of a given size (via a list comprehension), and uses it to generate
the first few Catalan numbers.  This list is then extended via lookup
in the Online Encyclopedia of Integer Sequences (OEIS) \cite{oeis}.

\begin{listing}
  \inputminted{text}{examples/catalan.disco}
  \caption{Counting trees}
  \label{lst:catalan}
\end{listing}

The first thing to note is that \disco has \emph{equirecursive}
algebraic types.  The \texttt{type} declaration defines the type
\texttt{BT} to be \emph{the same type as} \texttt{Unit + BT*BT} (\ie
the tagged union of the primitive one-element \texttt{Unit} type with
pairs of \texttt{BT} values).  This is a big departure from the
\emph{isorecursive} types of Haskell and OCaml, where
\emph{constructors} are required to explicitly ``roll'' and ``unroll''
values of recursive types.  We can see in the example that
\texttt{size} takes a value of type \texttt{BT} as input, but can
directly pattern-match on \texttt{left(unit)} and \texttt{right(l,r)}
without having to ``unfold'' or ``unroll'' it first.  Using
equirecursive types makes the implementation of the type system more
complex, but it is a very deliberate choice:
\begin{itemize}
\item There is less incidental complexity for students to stumble
  over.  In my experience, students learning Haskell are often
  confused by the idea of constructors and how to use them to create
  and pattern-match on data types.
\item \disco has no special syntax for declaring (recursive)
  sums-of-products; it simply has sum types, product types, and
  recursive type synonyms. Of course, it would be very tedious to
  write ``real'' programs in such a language---values of large sum
  types like \texttt{type T = A + B + C + ...} have to be written as
  \texttt{left(a)}, \texttt{right(left(b))},
  \texttt{right(right(left(c)))}, and so on. However, the sum types
  used as examples in a discrete mathematics class rarely have more than two
  or three summands, and working directly with primitive sum and
  product types helps students explicitly make connections to other
  things they have already seen, such as Cartesian product and
  disjoint union of sets.  It also reinforces the \emph{algebraic}
  nature of algebraic data types.
\end{itemize}

The \texttt{oeis} module is inessential, but can be a fun way for
students to explore integer sequences and the OEIS.  In addition to
\texttt{extendSequence}, the module also provides a
\texttt{lookupSequence} function, which returns the URL of the first
OEIS result, if there is any:
\begin{verbatim}
Disco> lookupSequence(catalan1)
right("https://oeis.org/A000108")
\end{verbatim}

The last things illustrated by this example are some facilities for
computing with collections.  The built-in \texttt{each} function is
like Haskell's \texttt{map}, but works for sets and multisets in
addition to lists.  \texttt{reduce} is like \texttt{foldr}, but again
working over sets and multisets in addition to lists.  In this case,
the \texttt{all} function is defined by first mapping a predicate over
\texttt{each} element of a list, then reducing the resulting list of
Booleans via logical conjunction.  (Putting twiddles (\verb|~|) in
place of arguments is the way to turn operators into standalone
functions, thus: \verb|~/\~|.)  Notice also that the \texttt{all}
function is polymorphic: \disco has support for standard parametric
polymorphism.

\subsection{Defining and testing bijections}

\pref{lst:bijection} shows part of another exercise I give to my
students, asking them to define the inverse of a given function and
use \disco to check that their inverse is correct.  This exercise makes
essential use of the testing facility we have already seen: if a
student defines a function which is not inverse to the given function,
\disco is usually able to quickly find a counterexample.  Running
this counterexample through the functions hopefully gives the student
some insight into why their function is not correct.  For example, if
we try (incorrectly) defining \texttt{g2(x) = x - 1/2}, \disco reports
\begin{verbatim}
  g2:
  - Certainly false: ∀x. f2(g2(x)) == x
    Counterexample:
      x = 1
\end{verbatim}

\begin{listing}
  \inputminted{text}{examples/bijection.disco}
  \caption{Defining and testing bijections}
  \label{lst:bijection}
\end{listing}

In this example we can also see more numeric types besides the natural
numbers.  \disco actually has four primitive numeric types:
\begin{itemize}
\item The natural numbers $\N = \{0, 1, 2, \dots\}$, which support
  addition and multiplication.
\item The integers $\Z = \{\dots, -2, -1, 0, 1, 2, \dots\}$, which
  besides addition and multiplication also support subtraction.
\item The \emph{fractional numbers} $\F = \{ a/b \mid a, b \in \N, b
  \neq 0 \}$, \ie nonnegative rationals, which besides addition and
  multiplication also support division.
\item The \emph{rational numbers} $\Q$, which support all four
  arithmetic operations.
\end{itemize}

\disco uses \emph{subtyping} to match standard mathematical practice.
For example, it is valid to pass a natural number value to a function
expecting an integer input.  Mathematicians (and students!) would find
it very strange and tedious if one were required to apply some sort of
coercion function to turn a natural number into an integer.

\begin{figure}[htp]
  \centering
  \includegraphics[width=0.3\textwidth]{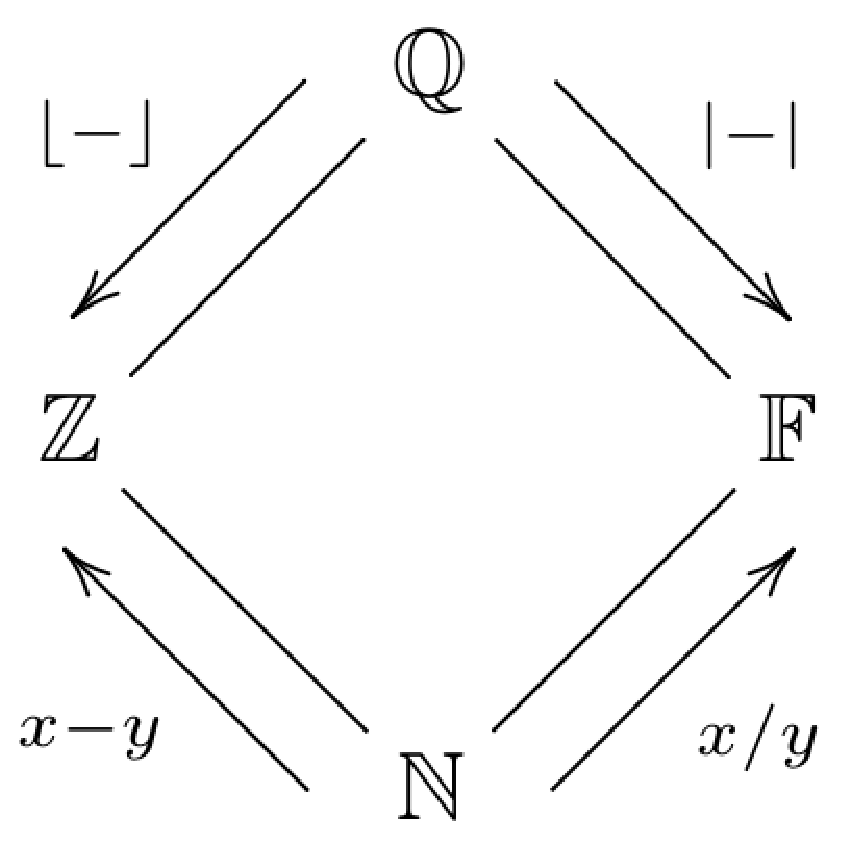}
  \caption{\disco's numeric type lattice}
  \label{fig:lattice}
\end{figure}

These four types naturally form a diamond-shaped lattice, as shown in
\pref{fig:lattice}.
$\N$ is a subtype of both $\Z$ and $\F$, which are in turn both
subtypes of $\Q$.  Moving up and left in the lattice (from $\N$ to
$\Z$, or $\F$ to $\Q$) corresponds to allowing subtraction; moving up
and right corresponds to allowing division.  Moving down and left can
be accomplished via a rounding operation such as floor or ceiling;
moving down and right can be accomplished via absolute
value. \pref{lst:subtype} demonstrates these ideas by requesting the
types of various expressions.  In the last example, in particular,
notice how \disco infers the type $\Q$ for the elements of the list,
since that is the only type that supports both negation and division.
\begin{listing}
\begin{minted}{text}
Disco> :type -3
-3 : ℤ
Disco> :type |-3|
abs(-3) : ℕ
Disco> :type 2/3
2 / 3 : 𝔽
Disco> :type -2/3
-2 / 3 : ℚ
Disco> :type floor(-2/3)
floor(-2 / 3) : ℤ
Disco> :type [1,2,3]
[1, 2, 3] : List(ℕ)
Disco> :type [1,-2,3/5]
[1, -2, 3 / 5] : List(ℚ)
\end{minted}
\caption{Numeric types and subtyping}
\label{lst:subtype}
\end{listing}

\disco has no floating-point type, because floating-point numbers are
\emph{the worst} \cite{goldberg1991every} and there is no particular
need for real numbers in a discrete mathematics course.


\section{Discussion and Future Work}
\label{sec:discussion}

\subsection{Syntax}
\label{sec:syntax}

For the most part, \disco tries to use syntax as close to standard
mathematical syntax as possible.  However, there are a few notable
cases where this was deemed impossible, typically because standard
mathematical syntax is particularly ambiguous or overloaded.  Thinking
explicitly about these cases is a worthwhile exercise, since they are
likely to be confusing to students anyway.

\begin{itemize}
\item Mathematicians are very fond of using vertical bars for multiple
  unrelated things, and \disco actually does well to allow them in
  many cases: absolute value, set cardinality, and the separator
  between expression and guards in a comprehension all can be written
  in \disco with vertical bars.  However, the ``is a divisor of''
  relation is also traditionally written with a vertical bar, as in
  $3 \mid 21$, but \disco does not support this notation.  Including
  it would make the grammar extremely ambiguous.  (And besides,
  Dijkstra would tell us that we should not use a visually symmetric
  symbol for a nonsymmetric relation!) Instead, \disco provides
  \texttt{divides} as an infix operator.  In my experience students
  have no problem remembering the difference.
\item In mathematics, the equality symbol $=$ is also typically
  overloaded to denote both definition (``\emph{let $x = 3$, and
    consider\dots}'') and equality testing (``\emph{if $x = 3$,
    then\dots}'').  \disco cannot use the same symbol for both, since
  otherwise it would be impossible to tell whether the user is writing
  a definition or entering a Boolean test to be evaluated.  This is
  confusing for students but it seems like it can't be helped, and in
  any case I would argue that trying to gloss over the difference is
  not really doing students any favors, but simply allowing them to
  persist in some fundamental misunderstandings.
\item \disco allows juxtaposition to denote both function application,
  as in \texttt{f(3)}, and multiplication, as in \texttt{2x}.  It uses
  a simple syntax-directed approach to tell them apart: if the
  expression on the left-hand side of a juxtaposition is a numeric
  literal, or a parenthesized expression with an operator, then it is
  interpreted as multiplication; otherwise it is interpreted as
  function application.  However, this does not always get it right,
  and there are times when an explicit multiplication operator must be
  written (one notable example is when scaling the output of a
  function call, as in $2f(x)$; in \disco this must currently be
  written \mintinline{text}{2*f(x)}, although this can hopefully be
  fixed).  It might be worth exploring a more type-directed approach,
  although that would be considerably more complex.  It seems like to
  really get this ``right'' requires general intelligence: for
  example, does the expression $f(x+2)$ denote multiplication or
  function application?  Are you sure?  How do you know?  What about
  in the expression $x(y+2)$?  Or how about ``Let $x$ be the function
  which doubles its argument, and consider $x(y+2)$ \dots''?
\end{itemize}

\subsection{Student experience}
\label{sec:students}

So far, I have used \disco in my discrete mathematics course twice, in
the spring semesters of 2022 and 2023.  Both courses had about 25
students, mostly sophomore computer science majors, with a few
mathematics majors in the mix as well.  In the spring of 2023 I asked
students a couple questions about Disco on the end-of-semester course
evaluation.  The results are shown below.  Although I did not get a
good response rate and the results have no statistical significance,
they are at least encouraging.  The few students who wrote optional
textual comments also had very positive things to say.

\newcolumntype{P}[1]{>{\centering\arraybackslash}p{#1}}

\begin{table}[]
\begin{tabular}{p{2in}P{0.65in}P{0.65in}P{0.65in}P{0.65in}P{0.65in}}
& Strongly Disagree & Disagree & Neutral & Agree & Strongly Agree \\
Using Disco helped me learn the mathematical ideas in this course better. & 0                 & 2        & 0                          & 7     & 7              \\
Learning Disco helped me improve my computer programming skills.          & 0                 & 0        & 4                          & 7     & 5
\end{tabular}
\end{table}


\subsection{Type system}
\label{sec:types}

\disco and its type system were designed to be intuitive for students
and to corresponding closely to mathematical practice, but this has
not always led to the simplest type system from an implementation
point of view!

\begin{itemize}
\item As previously mentioned, \disco has \emph{subtyping} in order to
  accommodate typical mathematical practice.  \disco's subtyping is
  \emph{structural}, meaning that we only really need concern
  ourselves with subtyping relationships between primitive types; a
  subtyping relation between complex types (for example, sum, product,
  or function types) can always ultimately be broken down into
  subtyping relations between simpler types.  Subtyping complicates
  the type system since, for example, when typechecking the
  application of a function to an argument, we cannot just check that
  the types match via unification, but we must instead emit a
  subtyping constraint which we check later.

\item \disco also has parametric polymorphism, since a language
  without polymorphism would not really give students a good idea of
  the expressive power of statically typed functional programming.  Of
  course, this means that typing constraints can involve unification
  variables as well as \emph{skolem variables} (when \emph{checking} a
  polymorphic type).

\item \disco's type system must actually support \emph{qualified}
  polymorphism (similar to Haskell's type classes, but with only a
  specific set of built-in classes) in order to be able to infer types
  in a setting where some types support certain operations (\eg
  subtraction or division) and some do not.  For example, what is the
  type of \mintinline{text}{\x. x - 2}?  Most generally, this function
  has a type like
  $\forall a.\; (\textit{sub}(a), \Z <: a) \Rightarrow a \to a$, that
  is, is a polymorphic function with type $a \to a$ for any type $a$
  which supports subtraction and has $\Z$ as a subtype, \ie either
  $\Z$ or $\Q$.  (As a nice exercise, you might like to convince
  yourself that none of $\Z \to \Z$, $\Z \to \Q$, $\Q \to \Z$, or $\Q
  \to \Q$ will work---some of them are invalid types for the function,
  and some of them, although valid, are not general enough.)

  Such types are currently only allowed internally, during type
  inference, but must be monomorphized away before showing types to
  users. This is sound, but can be rather confusing.  For example,
  \disco will report that the type of \mintinline{text}{\x. x - 2} is
  $\Z \to \Z$, but will also happily allow it to be applied to a
  fractional input such as $5/2$, which would be a type error if its
  most general type were really $\Z \to \Z$.
\begin{verbatim}
Disco> :type \x. x - 2
λx. x - 2 : ℤ → ℤ
Disco> (\x. x - 2) (5/2)
1/2
\end{verbatim}
  One interesting idea to improve the situation would be to show the
  user \emph{multiple} potential monomorphic instantiations of a
  general type scheme, something like this, perhaps:
\begin{verbatim}
Disco> :type \x. x - 2
λx. x - 2
  : ℤ → ℤ
  : ℚ → ℚ
\end{verbatim}
\item As mentioned before, \disco has equirecursive types.  The big
  wrinkle this adds to the type system is that simple structural
  equality (or unification) no longer suffices; when recursive type
  synonyms are involved, two types can be the same even though they
  look different.
\end{itemize}

The combination of qualified parametric polymorphism, subtyping, and
equirecursive types makes for an overall system which seems only
barely on the edge of tractability.  For the implementation of type
inference and checking I relied heavily on Traytel \etal
\cite{traytel2011extending} who describe the implementation of a
similar type system for Isabelle/HOL.  There are almost certainly
bugs, but overall I am fairly confident in the soundness of the type
system.

\subsection{Types vs sets}
\label{sec:typesvsets}

Axiomatic set theory is usually taken as the \emph{de facto}
foundation for mathematics.  On the other hand, in practice,
mathematicians usually behave more as if they were working in some
kind of type-theoretic foundation, which makes a statically typed
functional language a good match for mathematics as it is practiced
(for example, see HoTT \cite{hottbook} and Lean \cite{moura2021lean}).
However, one area where there seems to be a big mismatch is in the
distinction between \emph{types} and \emph{sets}.

To most mathematicians and every discrete mathematics textbook ever,
$\{2,4,7\}$ and $\N$ are both examples of \emph{sets}.  The former is
finite and the latter (countably) infinite, but they are both
fundamentally the same kind of thing, and it makes sense to talk about
(for example) their difference, $\N - \{2,4,7\}$, which is also a set.
In \disco, however, $\{2,4,7\}$ and $\N$ are very different things:
the former is a value of type \mintinline{text}{Set(N)}, whereas the
latter is a type, and $\N - \{2,4,7\}$ is so nonsensical that it is a
\emph{syntax} error!  One might reasonably wonder: why the mismatch?
Why not try to make \disco more closely align with common mathematical
practice, in accordance with \disco's stated philosophy?

Although conflating sets and types might be fine on a theoretical
level (at least, as long as one does not worry about deeper
foundational issues), on a practical level it introduces several
big problems:

\begin{itemize}
\item The ability to use arbitrary finite sets as types would lead to
  what is essentially a system of \emph{refinement types}.  Although
  this is well-studied and has many practical motivations, it quickly
  leads to undecidable typechecking, the need for tools like SMT
  solvers, and the requirement for users to provide annotations to
  help the system understand why a given type is valid.  For example,
  to typecheck %
  \verb|f : N -> {2,3,7}| would require somehow checking that for any
  natural number input, the function \verb|f| will always return
  either $2$, $3$, or $7$, which could depend on complex reasoning
  about the behavior of the function.  Calling out to an SMT solver in
  order to typecheck a teaching language to be used by students seems
  like a non-starter.
\item Conversely, the ability to use types as value-level sets
  introduces all sorts of difficulties, chief among which is the fact
  that most types correspond to \emph{infinite} sets.  Set values
  would have to be represented at runtime as some kind of abstract set
  expressions rather than simply as sets of values, and operations
  like membership checking become only \emph{semi}-decidable at best.
  What's more, these infinite set values would not really correspond
  to their supposed mathematical counterparts in some subtle ways.
  For example---as we often teach discrete mathematics students---the
  power set of the natural numbers is uncountable; but if \texttt{N}
  were usable as a value of type \texttt{Set(N)} in \disco, then
  \texttt{power(N)} would actually represent the set of
  \emph{computable} subsets of $\N$, which is countable!
\end{itemize}

The one slight blurring of categories which seems both feasible and
desirable would be the ability to use finite set values as domains for
$\forall$ and $\exists$ property quantifiers, so one could write, for
example, \mintinline{text}{forall x in [0..10]. x^2 <= 100}.  It seems
theoretically straightforward to incorporate such finite sets into the
existing machinery for property checking, though this has not been
done yet.

In any case, how should we present and explain the relevant
distinctions to students?  Honestly, I'm not entirely sure.  My best
approach at the moment revolves around two ideas:
\begin{itemize}
\item First, explain to students that \disco can only represent
  \emph{finite} sets.  This is easy enough to understand: if we
  allowed infinite sets, then certain operations might require
  infinitely long computations.
\item We can then explain that types can be thought of as a particular
  collection of ``ur-sets'' out of which we can build and carve out
  all other sets.  For particularly keen students, we can explain that
  types are sets with particularly nice structural properties.  For
  example, $\N$ is the unique set that includes $0$ and is closed
  under the successor operation; in contrast, there is no nice
  structural way to define $\{2,4,7\}$ besides just listing its
  elements.  These nice structural properties are precisely what
  enable decidable typechecking without having to resort to SMT
  solvers.
\end{itemize}

\subsection{Formal proofs}
\label{sec:formal-proof}

A discrete mathematics course often includes an introduction to proof
writing.  Currently, Disco only helps with this indirectly: it helps
students practice expressing themselves formally, and property-based
testing can give early feedback to see whether a particular
proposition is worth trying to prove.  Ideally, however, there would
be a way to use Disco more directly in constructing formal proofs, or
a way to integrate it with existing tools for doing so.

\subsection{Error messages}
\label{sec:errors}

When the \disco project first started, I had grand designs for the way
the system would interact with the user in the case of type errors
\cite{yorgey2018explaining}. Unfortunately, partly because I was
intimidated by my own grand designs, and partly because error messages
are hard, the system currently does not have very good error messages!
For example, here is a terribly uninformative one:
\begin{verbatim}
Disco> each(3, [1,2,3])
Error: the shape of two types does not match.
https://disco-lang.readthedocs.io/en/latest/reference/shape-mismatch.html
\end{verbatim}
In practice, I just tell students to ask me for help when they run
into errors they can't figure out, but this obviously limits wider
adoption.  Improving error messages will be another big focus for work
in the upcoming year.

\section{Acknowledgments}
\label{sec:acks}

\disco was originally born out of a conversation with Harley Eades at
TFPIE in 2016, and I'm very grateful to Harley for those initial
conversations and many good ideas.  I am also grateful to the many
people who have contributed to the \disco codebase over the years,
including my students Callahan Hirrel, Bosco Ndemeye, Sanjit
Kalapatapu, Jacob Hines, Eric Pinter, and Daniel Burnett, as well as
other collaborators including Shay Lewis, Ryan Yates, Tristan de
Cacqueray, and Chris Smith.

\bibliographystyle{eptcsalpha}
\bibliography{references}
\end{document}